\documentclass{cccg10}
\pdfoutput=1
\usepackage{graphicx,amssymb,amsmath,hyperref}

\graphicspath{{colorfigs/}}
\title{Regular Labelings and Geometric Structures}

\author{David Eppstein\thanks{Computer Science Department,
University of California, Irvine, {\tt eppstein@uci.edu}}}
        
\index{Eppstein, David}

\begin{document}
\thispagestyle{empty}
\maketitle

\begin{abstract}
Three types of geometric structure---grid triangulations, rectangular subdivisions, and orthogonal polyhedra---can each be described combinatorially by a \emph{regular labeling}: an assignment of colors and orientations to the edges of an associated maximal or near-maximal planar graph. We briefly survey the connections and analogies between these three kinds of labelings, and their uses in designing efficient geometric algorithms.
\end{abstract}

\section{Introduction}
In a remarkable 1990 paper, Walter Schnyder described an efficient algorithm for drawing any maximal planar graph, using straight line edges, in an integer grid of linear dimensions~\cite{Sch-SODA-90}. A key tool in Schnyder's result is a \emph{regular labeling} of the edges of the graph by colors and orientations, satisfying local constraints on how the edges incident to individual vertices can be labeled. Soon after Schnyder's work, Kant and He~\cite{He-SJC-93,KanHe-TCS-97} showed that \emph{rectangular partitions} of a rectangle into smaller rectangles could also be described as a regular labeling, with different local constraints. More recently with Elena Mumford we showed that another kind of regular labeling describes \emph{orthogonal polyhedra} in which three perpendicular edges meet at each vertex~\cite{EppMum-SoCG-10}. Each of these labelings may be used algorithmically to efficiently construct the structures it describes.

There are many strong analogies between these three types of labeling: each involves a maximal or near-maximal planar graph with a designated outer face and can be described as a coloring and orientation of the graph's edges. In each case local constraints at the vertices automatically ensure the acyclicity of certain subgraphs derived from the labeling. And, in each case, the set of labelings of a fixed graph forms a distributive lattice in which adjacent lattice elements are related by local twisting operations. However the reasons that these three kinds of geometric object should be described combinatorially in such similar ways to each other are not well understood.

In this paper we survey these three types of regular labeling, their properties, and their applications, with the hope that bringing this material together will bring to light further connections between these problems.

\section{Grid triangulations}
According to F\'ary's theorem~\cite{Far-ASMS-48,Ste-PotAMS-51,Wag-JGM-36}, any crossing-free drawing of a graph in the plane with curved edges can be straightened to a combinatorially equivalent drawing in which all edges are straight line segments. More strongly, it is possible to find a straight line drawing in which all vertices are placed in an integer grid of linear dimensions~\cite{ChrPay-IPL-95,FraPacPol-Comb-90,Kan-Algo-96,Sch-SODA-90}. The proof of this result that concerns us is the one by Schnyder~\cite{Sch-SODA-90} based on the idea of a regular labeling.

Let $G$ be a maximal planar graph; fix a planar embedding of $G$ by specifying an outer face and its orientation. We define a \emph{Schnyder labeling} of $G$ to be an assignment of one of the three colors \{red, blue, green\} to each edge other than the three outer edges, and an orientation to each edge, with the following properties:
\begin{itemize}
\item At each of the three outer vertices of $G$, all edges are oriented inwards, and all have the same color. The edges at the three vertices, in clockwise order, have the three colors red, blue, and green.
\item Each non-outer vertex of $G$ has exactly one outgoing edge of each color, in the cyclic order red, blue, and green. The incoming edges between any two of these outgoing edges all have the same color, which must be different from the color of the two outgoing edges that surround them.
\end{itemize}

\begin{figure*}[t]
\centering\includegraphics[scale=0.85]{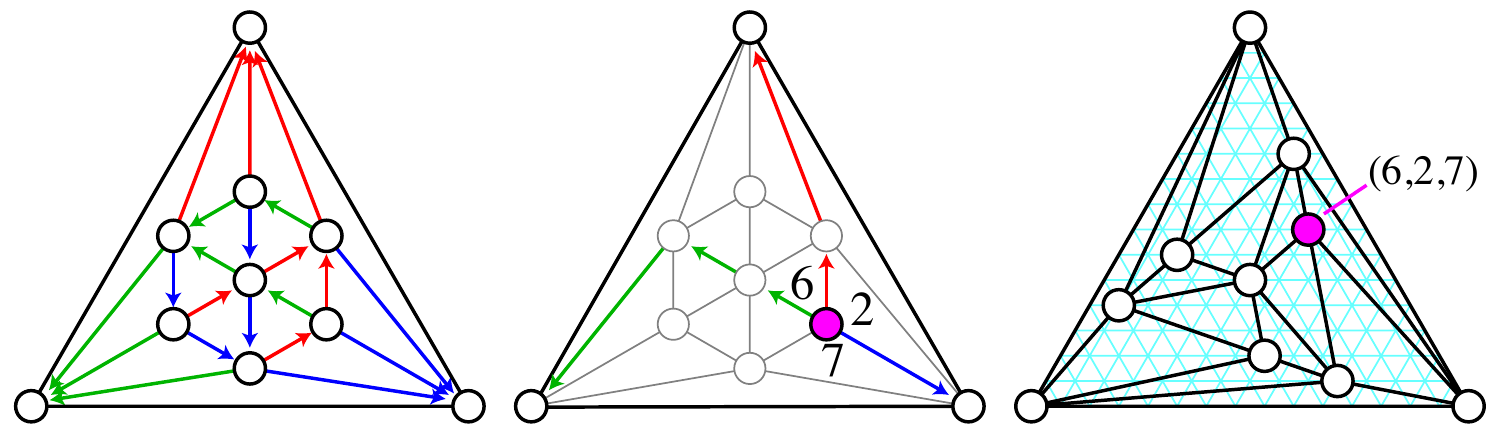}
\caption{A Schnyder labeling (left), the counts of faces in the three regions formed by the monochromatic outgoing paths from a vertex (center), and the grid embedding formed by using face counts as barycentric coordinates (right).}
\label{fig:SchnyderEmbedding}
\end{figure*}

Figure~\ref{fig:SchnyderEmbedding} (left) shows an example.
In a Schnyder labeling, each monochromatic subgraph must be acyclic. For, suppose to the contrary that there is a cycle; without loss of generality, by symmetry, we may suppose this cycle to be red and embedded in the plane with a clockwise orientation. Then at each vertex of the cycle there would be a blue edge oriented into the cycle. Therefore, the path of blue edges extending from one of these vertices cannot escape the red cycle and must eventually repeat a vertex, forming a smaller blue cycle. But the same argument would show that this blue cycle contains a smaller green or red cycle, forming an infinite regress; this contradiction completes the proof.  A stronger form of acyclicity can be proven by a similar argument: every graph formed by choosing two of the three color classes, and reversing the edges in one of these two classes, is  an \emph{$st$-planar graph}, a plane graph oriented acyclically with a single source and a single sink which are both on the outer face of their embedding. As an acyclic graph with outdegree one and with only one sink vertex (one of the three vertices of the outer face), each monochromatic graph must be a tree, and these three trees form a structure known as a \emph{realizer} or a \emph{Schnyder wood}. A Schnyder labeling of any maximal planar graph with designated outer face can be constructed in linear time by contracting one vertex into the outer face, recursively constructing a labeling of the remaining graph, undoing the contraction, and performing local adjustments to the labeling.

Each internal vertex of $G$ has three monochromatic paths leading to the three external vertices. These paths cannot cross, as a crossing would form a cycle in one of the three bichromatic subgraphs, so they partition the $2n-3$ internal faces of $G$ into three contiguous regions (Figure~\ref{fig:SchnyderEmbedding}, center). We may use the numbers of faces in each region as \emph{barycentric coordinates} to place each vertex into an equilateral triangle of height $2n-3$; that is, if vertex $v$ has $k$ faces in the region bounded by the blue and green paths ($k=7$ in the figure) then it goes $k$ units away from the side of the triangle opposite the red vertex. The three outer vertices are placed at the vertices of the equilateral triangle. With this placement, each triangle of $G$ can be inscribed in an equilateral triangle that is rotated $180^\circ$ relative to the outer face. Every face of $G$ has a consistent orientation, so there can be no crossings and the placement forms a straight-line embedding (Figure~\ref{fig:SchnyderEmbedding}, right). Using two  face counts as Cartesian coordinates and ignoring the third affinely transforms this embedding into an integer grid.

One may also count vertices within the three regions formed by the monochromatic paths from $v$, instead of counting faces. Vertices on the clockwise boundary of the region are included within the region, and vertices on the counterclockwise boundary are excluded. In this way, the three vertex counts add to $n-1$ and can again be used as barycentric or Cartesian coordinates, leading to an embedding that fits within an $(n-2)\times (n-2)$ grid and takes time $O(n)$ to construct~\cite{Sch-SODA-90}. Although there have been some technical improvements to the grid size using the same approach~\cite{BonFelMos-Algo-07,ZhaHe-WADS-03}, this remains the record for the smallest grid known to contain any $n$-vertex planar graph~\cite{BonFelMos-Algo-07}. Dhandapani~\cite{Dha-DCG-10} instead assigns positive real weights to the faces of the drawing, and uses the sums of the weights in each of the three regions as barycentric coordinates. By applying a version of the Brouwer fixed point theorem to this system of weights, he shows that any maximal planar graph has a straight-line drawing in which each pair of vertices is connected by a distance-decreasing path.

If a triangle in a Schnyder labeling (not necessarily a face) has its edges labeled with all three colors, then it must be cyclically oriented. In this case it is possible to form a different Schnyder labeling by reversing the edges of the triangle and cyclically permuting the colors of its edges and the edges inside it. This operation may also be viewed as twisting one of the three monochromatic paths extending from each of the triangle's vertices either one unit clockwise or one unit counterclockwise. The graph with one vertex per Schnyder labeling of a graph $G$ and one edge per twist operation, directed from counterclockwise to clockwise, is acyclic, and defines a distributive lattice structure on the set of Schnyder labelings of~$G$~\cite{Bre-00,Fel-EJC-04,Oss-94}.

This distributive lattice can be described in a different way, in terms of fixed-outdegree orientations of a planar graph. The colors of the edges in a Schnyder labeling can be recovered by propagating colors from neighbor to neighbor once the orientations are known, so a Schnyder labeling can equivalently be described as an orientation of $G$ in which each internal vertex has outdegree three and each external vertex has outdegree zero~\cite{FraMen-DM-01,Fel-EJC-04}. The orientations of $G$ satisfying these constraints, as more generally with the orientations of a planar graph with fixed outdegrees at each vertex, form a distributive lattice, the same one generated by local twists~\cite{Fel-EJC-04}.

Schnyder also used his labelings to characterize planar graphs as the graphs whose incidence order is three-dimensional~\cite{Sch-Ord-89}. Schnyder labelings can be generalized to 3-connected planar graphs that are not necessarily triangulations~\cite{Fel-Ord-01,Fel-Ord-03,Fel-EJC-04}, and in this generalized form they correspond to embeddings of the graphs onto three-dimensional orthogonal surfaces in which all edges follow geodesics of the surface~\cite{Fel-Ord-03,Fel-EJC-04,Mil-DM-02}.

\section{Rectangular partitions}

\emph{Rectangular partitions}, partitions of a rectangle into smaller rectangles, have important applications in the design of rectangular cartograms (maps that distort geographic regions into rectangles in order to visualize numerical information as the areas of the distorted regions)~\cite{Rai-GR-34}, treemaps (visualizations of hierarchies as recursive partitions of a root rectangle into smaller rectangles)~\cite{BruHuiWij-DV-00,JohSch-Vis-91}, arrangements of rooms in the architectural design of buildings~\cite{EarMar-AoGT-79,Rin-EPB-88}, clustering and partitioning of numerical data arrays~\cite{MutPooSue-ICDT-99}, and the layout of functional units in VLSI circuits~\cite{LiaLuYen-Algs-03}.

The \emph{extended dual} graph of a rectangular partition is a planar graph with one vertex for each rectangle in the partition, one vertex for each side of the larger rectangle that is being partitioned, and an edge between two vertices if the rectangles or sides they correspond to are incident~\cite{KozKin-Nw-85}. When three rectangles meet at each internal vertex, this graph is nearly maximal planar, except that its outer face is formed by the four dual vertices corresponding to the sides of the outer rectangle, and is therefore a quadrilateral rather than a triangle. The extended dual can have no separating triangles, because a separating triangle would correspond to a region surrounded completely by only three rectangles, a geometric impossibility.

Each edge in the extended dual other than the edges between the four outside faces corresponds to an adjacency along a horizontal or vertical segment of the partition, and we may color the dual edge respectively red or blue according to the orientation of this adjacency, and orient it from the upper feature to the lower feature or from the left feature to the right feature. The result is a regular labeling~\cite{He-SJC-93,KanHe-TCS-97}; to distinguish it from the other regular labelings we survey, we call it a \emph{rectangular labeling}. It has the following defining properties:
\begin{itemize}
\item Each edge except for the four outer edges is colored red or blue and oriented.
\item Each outer vertex has edges of only one color and orientation.
\item Each interior vertex has edges of all four colors and orientations.
\item At each interior vertex, the edges with the same color and orientation are all contiguous, and these contiguous groups of edges occur in the cyclic order in-blue, in-red, out-blue, out-red.
\end{itemize}
Conversely, if $G$ is maximal planar except for a quadrilateral outer face, has no separating triangles, and is colored according to these conditions, then it forms the extended dual of a rectangular partition: the combinatorial structure of the partition may be read off directly from the coloring and orientation, and each line segment in the partition may be assigned as its $x$ or $y$ coordinate the length of the longest blue or red path whose final edge crosses the line segment. The consistency of this assignment follows from the $st$-planarity of the monochromatic subgraphs of the labeling, which can be proven using an argument based on an infinite regress of nested cycles, similarly to the proof of acyclicity of monochromatic subgraphs of Schnyder labelings.

Every graph $G$ that is maximal planar except for a quadrilateral outer face and has no separating triangles is an extended dual of a rectangular partition. A rectangular labeling for $G$ can be found in linear time by contracting an edge of $G$, recursively labeling the resulting smaller graph, uncontracting the edge, and adjusting the labeling~\cite{KanHe-TCS-97}. Thus, rectangular partitions with any realizable pattern of adjacencies between the rectangles can be found in linear time.
For closely related combinatorial representations of rectangular partitions, and an earlier linear-time algorithm for constructing a rectangular partition for a given dual graph, see also~\cite{BhaSah-Algo-88,TamTol-Allerton-89}.

\begin{figure}[t]
\centering\includegraphics[scale=0.5]{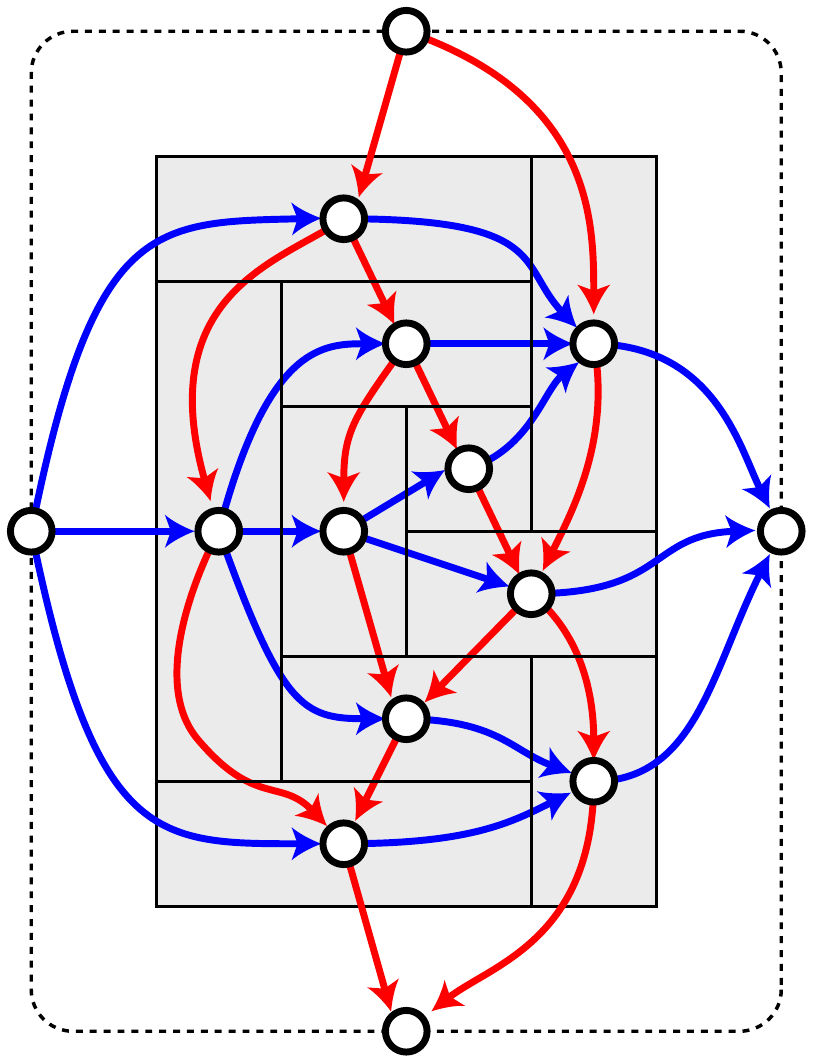}
\caption{A rectangular partition and its extended dual.}
\label{fig:RectangularDual}
\end{figure}

If $Q$ is a quadrilateral in a rectangular labeling that has its four edges colored with alternating colors, then one may obtain another valid rectangular labeling by changing the colors of all the edges interior to $Q$ and reversing the direction of one of the two sets of colored edges within $Q$; as with Schnyder labelings, this kind of move can be interpreted as twisting the boundaries between the four groups of edges at each of the four vertices of $Q$. All rectangular labelings of the same extended dual graph may be related to each other by sequences of twists, and the set of rectangular labelings of a given extended dual graph may be given the structure of a distributive lattice in which the twist operations connect pairs of labelings that form covering pairs in the lattice~\cite{Fus-GD-05,Fus-DM-09,TanChe-ISCAS-90}.

As with Schnyder labelings, this lattice also describes the fixed-outdegree orientations of a planar graph, the bipartite graph formed by the vertices and triangles of $G$. Each interior vertex of $G$ has four outgoing edges in this bipartite graph (representing the four breaks between groups of edges at that vertex), each triangle has one outgoing edge (to the vertex of $G$ incident to two edges of the triangle with the same color), and the four external vertices have no outgoing edges.

Because the set of rectangular labelings of a graph $G$ forms a distributive lattice, Birkhoff's representation theorem may be used to represent this set of labelings compactly, as the family of lower sets of a partial order~\cite{Bir-DMJ-37}. The elements of the partial order can be identified with pairs $(q,i)$ where $q$ is a quadrilateral of $G$ and $i$ is the number of times $q$ has been twisted relative to the bottom element of the lattice. This partial order can be described by a DAG with $O(n^2)$ vertices and edges, which can be constructed from $G$ in quadratic time~\cite{EppMum-WADS-09,EppMumSpe-SoCG-09}. With this compact representation of the space of rectangular labelings in hand, all labelings may be generated quickly. This representation has also been used in fixed-parameter tractable algorithms for finding area-universal rectangular partitions (that is, partitions that support all possible assignments of positive areas to each of the rectangles)~\cite{EppMumSpe-SoCG-09} and in polynomial time algorithms for finding rectangular partitions that obey constraints on the orientations of some of the boundaries between rectangles~\cite{EppMum-WADS-09}.

\section{Orthogonal polyhedra}

\begin{figure}[t]
\centering\includegraphics[height=1.75in]{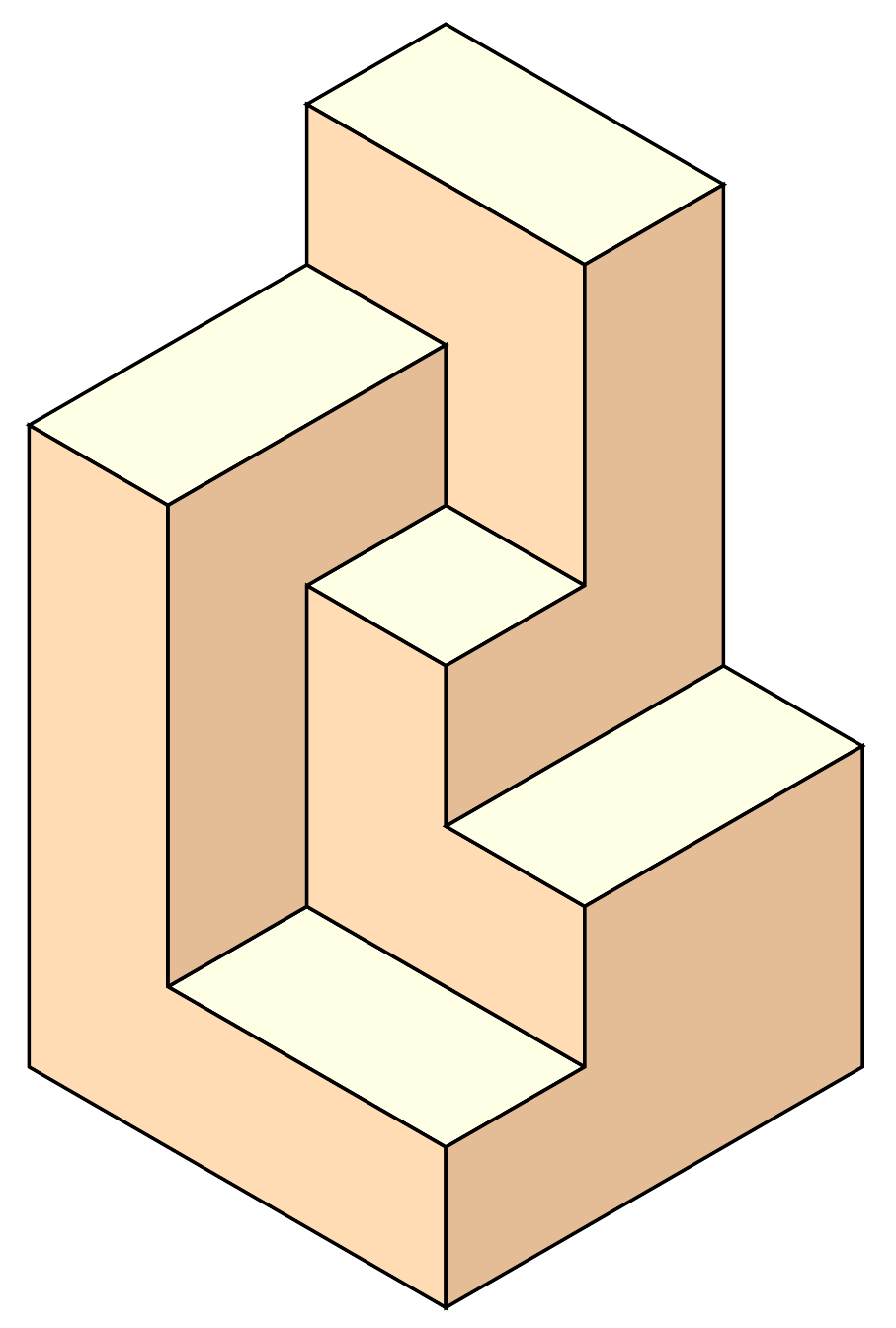}
\quad\includegraphics[height=1.95in]{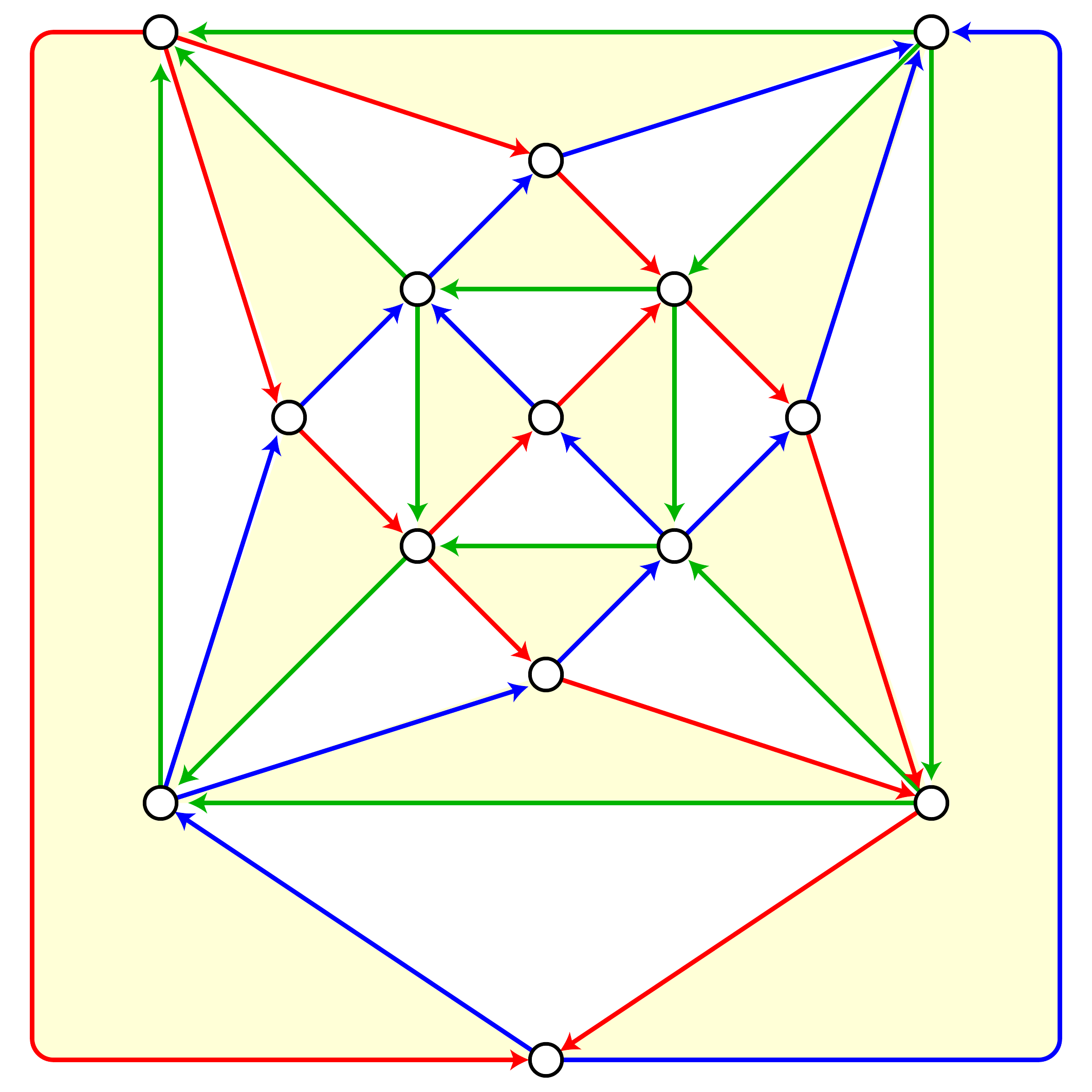}
\caption{The axonometric projection of a corner polyhedron, and its dual polyhedral labeling. The three outer vertices of the labeling correspond to the three back faces of the polyhedron, which are not shown in the figure.}
\label{fig:corner}
\end{figure}

Eppstein and Mumford~\cite{EppMum-SoCG-10} define a \emph{simple orthogonal polyhedron} to be a polyhedron with the topology of a sphere, with simply-connected faces, in which three mutually perpendicular edges meet at each vertex. They characterize the skeletons of these polyhedra graph-theoretically as the 3-regular 2-connected bipartite planar graphs in which the removal of any two vertices leaves at most two connected components. In this characterization, a central role is played by a more restricted class of polyhedra, the \emph{corner polyhedra}: these are the simple orthogonal polyhedra in which all but three faces are oriented in a positive direction. Corner polyhedra are closely related to solid Young diagrams and plane partitions~\cite{CohLarPro-NYJM-98}. Figure~\ref{fig:corner} (left) depicts an example, a corner polyhedron that is combinatorially equivalent to the truncated octahedron or permutohedron.

The dual graph of a corner polyhedron is maximal planar, with a distinguished outer triangle that corresponds to the three back faces of the polyhedron. Since every face of the polyhedron has an even number of edges, its underlying graph is bipartite and its dual graph is Eulerian. The bipartition of the underlying graph of the polyhedron dualizes to a 2-coloring of the triangles of the dual graph. A single dual graph may correspond to more than one corner polyhedron; for instance, the mirror reversal of the polyhedron depicted in Figure~\ref{fig:corner} is again a corner polyhedron, with an isomorphic dual graph but with a different choice of which edges are convex and which are concave. We may distinguish these representations combinatorially by another form of regular labeling, which we call here a \emph{polyhedral labeling}, in which we color the edges of the dual graph red, blue, or green according to which of the three coordinate axes the corresponding edges of the polyhedron are parallel to, and we orient each dual edge according to the coordinate ordering of the two faces corresponding to its endpoints (Figure~\ref{fig:corner}, right). This labeling has the following properties:
\begin{itemize}
\item Each triangle has edges of all three colors, and each vertex has edges of two colors.
\item At the outer three vertices, the orientations of the edges are in strict alternation.
\item At each inner vertex, the orientations of the edges alternate with exactly two breaks in the alternation, both of which are in triangles with the same color as the outer triangle.
\end{itemize}
If a labeling obeys these properties then, just as with a Schnyder labeling, the bichromatic subgraphs formed by choosing any two color classes of edges and reversing the edges in one of the two classes are necessarily $st$-planar. These $st$-planar orientations can be used to construct a corner polyhedron dual to a given Eulerian maximal planar graph, by using $st$-numberings of the three bichromatic graphs (an assignment of numerical values to vertices in which every edge is oriented from a smaller value to a larger one) as coordinates of the three parallel classes of face planes of the polyhedron. Thus, every polyhedral labeling uniquely encodes the orientations of edges in a corner polyhedron~\cite{EppMum-SoCG-10}.

If an Eulerian maximal planar graph is $4$-vertex-connected, it has at least one polyhedral labeling. It is not possible to prove this by an induction using edge contractions, as was done for Schnyder labelings and rectangular labelings: a single edge contraction does not preserve the Eulerian character of the graph. However, a more complex set of decomposition operations can reduce any Eulerian maximal planar graph to one of two simple base cases. Both base cases have polyhedral labelings, and the existence of a polyhedral labeling is preserved by the decomposition operations. Using these operations, it is possible to find a polyhedral labeling of any $4$-vertex-connected Eulerian maximal planar graph in linear randomized expected time or near-linear worst-case deterministic time. This construction forms a building block for algorithms that can construct more general orthogonal polyhedra from their graphs in the same time bounds~\cite{EppMum-SoCG-10}.

The family of polyhedral labelings on a graph $G$ can equivalently be described as orientations of an associated planar graph that connects the vertices of $G$ to the triangles that have the same color as the outer triangle. Each interior vertex of $G$ has two outgoing edges (the two triangles at which the orientation of edges in the labeling does not alternate), each triangle has one outgoing edge, and the three exterior vertices have no outgoing edges. Because polyhedral labelings correspond in this way to fixed-outdegree orientations on a planar graph, they again form a distributive lattice.

If an Eulerian maximal planar graph is not $4$-connected, it will still have a polyhedral labeling if and only if all separating triangles have the opposite color from the outer face when their interiors are removed and the triangular faces of the remaining graph are 2-colored. Thus, as in the case of rectangular labelings, separating triangles are central to the characterization of graphs with polyhedral labelings.

\section{Conclusions}

We have described three types of geometric object that have natural encodings as regular labelings of a maximal or near-maximal planar graph. Although much about these labelings is known, there are still many questions remaining:
\begin{itemize}
\item As with rectangular labelings, the distributive lattices of Schnyder labelings and polyhedral labelings admit a compact representation as the family of lower sets of a DAG of polynomial size. What are the algorithmic applications of this representation?
\item How can we recognize additional examples of this type? For instance, partitions of an equilateral triangle into smaller equilateral triangles seem to be closely related to 3-edge-colorings of a dual quadrilateral mesh with a hexagonal outer face in which the bichromatic subgraphs are $st$-planar; do they have a similar theory?
\item What unifying principle explains the close relation among these three different labelings? The equivalence to fixed-outdegree orientations of planar graphs explains some of their common features including the distributive lattices on their state spaces, and provides connections with plane partitions, domino tilings, and some other less-geometric distributive lattices derived from planar graphs~\cite{Fel-EJC-04,Pro-93},
but it does not explain why such distinct geometric objects should have such similar combinatorial representations.
\end{itemize}

\small\paragraph{Acknowledgements}
This research was supported in part by the National Science
Foundation under grant 0830403, and by the
Office of Naval Research under MURI grant N00014-08-1-1015. The author is grateful to Maarten L\"offler, Rodrigo Silveira, and Bettina Speckmann for helpful suggestions on a draft of this survey.

 \raggedright
\bibliographystyle{abuser}
\bibliography{reglabel}

\end{document}